\documentclass[acmsmall]{acmart}

\usepackage{amsmath,amsthm}

\usepackage{graphicx,tikz}

\usepackage{caption,subcaption}
\usepackage{enumitem}
\setlist[itemize]{leftmargin=5.5mm}

\usepackage{listings}
\usepackage{hyperref}

\newcommand{\remove}[1]{}

\author{Vitaly Aksenov}
\affiliation{ITMO University and IST Austria}
\email{aksenov@corp.ifmo.ru}

\author{Dan Alistarh}
\affiliation{IST Austria}
\email{dan.alistarh@ist.ac.at}

\author{Petr Kuznetsov}
\affiliation{LCTI, T\'el\'ecom ParisTech, Universit\'e Paris-Saclay}
\email{petr.kuznetsov@telecom-paristech.fr}

\begin{document}

\onecolumn

\title{Performance Prediction for Coarse-Grained Locking }

\begin{abstract}
A standard design pattern found in many concurrent data
structures, such as hash tables or ordered containers,
is an alternation of parallelizable sections that incur no data
conflicts and critical sections that must run sequentially and
are protected with locks.    
A lock can be viewed as a \emph{queue} that arbitrates the order
in which the critical sections are executed,   
and a natural question is whether we can use \emph{stochastic analysis}
to predict the resulting throughput.  
As a preliminary evidence to the affirmative, we describe a simple
model that can be used to predict the throughput of
\emph{coarse-grained} lock-based algorithms.
We show that our model works well for CLH lock, and
we expect it to work for other popular lock designs
such as TTAS, MCS, etc.
\end{abstract}

\maketitle




\section{Abstract coarse-grained synchronization}
\label{sec:problem}

Conventionally, the performance of a concurrent data
structure is evaluated via experiments, and 
it is notoriously difficult to account for all significant
experimental parameters so that the outcomes are meaningful.
Our motivation here is to complement experimental evaluation with an
analytical model that can be used to \emph{predict} the performance rather
than measure it.    
As a first step towards this goal,
in this work, we attempt to predict the throughput of a class of
algorithms that use \emph{coarse-grained} synchronization.   

Consider a concurrent system with $N$ processes that obey  
the following simple \emph{uniform} scheduler:
at every time step, each process performs a step of computation.
This scheduler,  resembling the well-known PRAM model~\cite{jaja1992introduction}, 
appears to be a reasonable approximation of a real-life concurrent system.
Suppose 
that the processes share a data structure exporting a single
\texttt{operation}$()$.
If the operation induces a work of size $P$ and incurs no synchronization,
the resulting throughput is $N \cdot \alpha / P$ operations in a unit of time:
each process performs $\alpha / P$ operations in
a unit of time, where $\alpha$ 
indicates the amount of work that can be performed by one process
in a unit of time. 
One way to evaluate the constant $\alpha$ experimentally is to count
the total number $F$ of operations, each of work $P$,   
completed by $N$ processes in time $T$.
Then we get $\alpha=F/NP$.
The longer is $T$, the more  accurate is the estimation of $\alpha$.

Now suppose that, additionally, the operation performed by each process 
contains a \emph{critical section} of size $C$.
In the operation, described in Figure~\ref{fig:operation},    
every process takes a global lock,
performs the critical section of size $C$,
releases the lock and, finally, performs
the \emph{parallel section} of size $P$.
\begin{figure}
\begin{lstlisting}
operation():
  lock.lock() 
  for i in 1..C:  |\label{line:crit1}|    
    nop |\label{line:crit2}|
  lock.unlock()
  for i in 1..P: |\label{line:par1}|
    nop |\label{line:par2}|
\end{lstlisting}
\caption{The coarse-grained operation}
\label{fig:operation}
\end{figure}

Here, as a unit of work, we take the number of CPU cycles spent during
one iteration of the loop in Lines~\ref{line:crit1}-\ref{line:crit2}
or~\ref{line:par1}-\ref{line:par2}.
%
The iteration consists of a \texttt{nop} instruction,
an increment of a local variable and a conditional jump, giving us, approximately,
\emph{four} CPU cycles in total.

\section{Model assumptions}

Below we list basic assumptions on the abstract machine used for our
analytical throughput prediction.

First, we assume that coherence of caches is maintained by
a variant of MESI protocol~\cite{papamarcos1984low}.
%
%
Each cache line can be in one of four states: \texttt{Modified} (\texttt{M}),
\texttt{Exclusive} (\texttt{E}), \texttt{Shared} (\texttt{S}) and \texttt{Invalid} (\texttt{I}).                        
MESI regulates transitions between states of a cache line and responses
depending on the request (read or write) to the cache line by a process
or on the request to the memory bus.
The important transitions for us are:
(1)~upon reading, the state of the cache line changes from any state to \texttt{S},
and, if the state was \texttt{I}, then a \emph{read request} is sent to the bus;
(2)~upon writing, the state of the cache line becomes \texttt{M},
and, if the state was \texttt{S} or \texttt{I}, an \emph{invalidation request} is sent to the bus. 

We assume that the caches are \emph{symmetric}:
for each MESI state $st$, there exist two constants $R_{st}$ and $W_{st}$
such that any read from any cache line with status $st$ takes $R_{st}$ work and
any write to a cache line with status $st$ takes $W_{st}$ work. 
%
\remove{
In our experiments, we use a 4-processor Intel Xeon E7-4870
2.4 GHz server with 10 threads per processor.
Compared to our abstract machine it has three level non-symmetric cache.
As a result,  accesses to cache lines in different sockets
can cost differently.
}
David et al.~\cite{david2013everything} showed that for an Intel Xeon
machine (similar to the one we use in our experimental validation
below), 
given the relative location of a cache line with respect to the process
(whether they are located on the same socket or not), the
following hypotheses hold:
(1)~writes induce the same work, regardless of the state of the cache line;
(2)~\texttt{swap}s, not concurrent with other \texttt{swap}s, induce the same work as writes.
Therefore, we assume 
that (1)~$W = W_M = W_E = W_S = W_I$ and
(2)~any contention-free \texttt{swap} induces a work of size $W$.

\section{CLH Lock}
Multiple lock implementations have been previously proposed,
from simple spinlocks and TTAS
to more advanced MCS~\cite{mellor1991algorithms} and CLH~\cite{craig1993building}.
For our analysis, we choose CLH, as the simplest lock among those
considered to be efficient.
\remove{
An important factor for the throughput is a choice
of the lock.
There are a lot of lock implementations:
spin lock, test\&test\&set, MCS~\cite{mellor1991algorithms}, CLH~\cite{craig1993building}, etc.
For simpicity, we take CLH lock: it is efficient in practice
and it is easier to analyse than other locks.
}
%
In Figure~\ref{fig:expand:operation}, we inline \texttt{lock} and \texttt{unlock} calls to CLH lock
in our abstract coarse-grained operation. 

\begin{figure}
\begin{lstlisting}
class Node:
  bool locked

Node head = new Node()  // global
Node my_node            // per process
   my_node.locked $\leftarrow$ true

operation():
  Node next $\leftarrow$ swap(&head, my_node)  // $W$ or $X$    |\label{line:swap}|
  while (next.locked) {}             // $R_I$ or $2 \cdot R_I$  |\label{line:wait}|  
  for i in 1..C:                     // $C$    |\label{line:critical:1}|
    nop                                        |\label{line:critical:2}|
  my_node.locked $\leftarrow$ false            // $W$    |\label{line:small:1}|      
  my_node $\leftarrow$ next                       |\label{line:small:2}|
  my_node.locked $\leftarrow$ true             // $W$  |\label{line:set}|
  for i in 1..P:                     // $P$    |\label{line:parallel:1}|
    nop                                        |\label{line:parallel:2}|
\end{lstlisting}
\caption{The coarse-grained operation with
inlined \texttt{lock} and \texttt{unlock} functions}
\label{fig:expand:operation}
\end{figure}

\subsection{Cost of an operation}
Let us zoom into what happens during the execution of the operation.

Note that at the beginning of an operation (unless it is the very first invocation),
\texttt{my\_node.locked} is loaded into the cache and the corresponding
cache line is in state \texttt{M}, because of the set in Line~\ref{line:set}
during the previous operation by the same process.


\noindent
(1)~The operation starts with \texttt{swap} (Line~\ref{line:swap})
that induces a work of size $W$, if not concurrent with other \texttt{swap}s,
and a work of size at most $X$, otherwise.
%

\noindent
(2)~In Line~\ref{line:wait}, the algorithm loops on a field \texttt{next.locked}.
During this loop one or two cache misses happens.

One cache miss can happen at the first iteration of the loop
if the read of \texttt{locked} returns \texttt{true}.
The last process that grabbed the lock
already invalidated this
cache line in Line~\ref{line:set} during its penultimate operation.
MESI reloads the cache line and changes its state from \texttt{I}
(or none if it was not loaded previously) to \texttt{S}.

The other cache miss happens in every execution when
the operation reads \texttt{next.locked} and gets \texttt{false}.
In this case, the cache line was invalidated in Line~\ref{line:small:1}
during the last operation of the last process that grabbed the lock.
MESI reloads the cache line and changes its state from \texttt{I} (or none) to \texttt{S}.

Each of the described cache misses induces the work of size $R_I$.
Thus, the work induced in Line~\ref{line:wait} is of size of $R_I$ (if only the
second miss happens) or $2 \cdot R_I$ (if both misses happen).

\noindent
(3)~In Lines~\ref{line:critical:1}-\ref{line:critical:2},  the critical section
with work of size $C$ is performed.

\noindent
(4)~In Line~\ref{line:small:1}, \texttt{my\_node.locked} is set to \texttt{false}.
There are two cases: if \texttt{my\_node.locked} is not yet loaded by any other process
in Line~\ref{line:wait} then the state remains \texttt{M};
otherwise, MESI changes the state from \texttt{S}
to \texttt{M} and sends a signal to invalidate this cache line.
In both cases, the induced work is of size $W$.
%

\noindent
(5)~In Line~\ref{line:small:2}, the operation performs an assignment on
local variables, without contributing to the total work.

\noindent
(6)~In Line~\ref{line:set}, \texttt{my\_node.locked} is set to
\texttt{true}. 
From the end of the while loop at Line~\ref{line:wait} the corresponding
cache line is in state \texttt{S}.
MESI changes the state to \texttt{M} and sends a signal to
invalidate this cache line inducing work of size $W$.

\noindent
(7)~In Lines~\ref{line:parallel:1}-\ref{line:parallel:2},
the parallel work of size $P$ is performed.

%

%



\subsection{Evaluating throughput}
%
%

To evaluate the throughput of the resulting program under the uniform scheduler,  
take a closer look on how $N$ processes continuously perform the
operation from Figure~\ref{fig:expand:operation}.
%

Process $1$ executes:
its first swap (taking at most $X$ units);
the critical section (blue, Lines~\ref{line:wait}-\ref{line:small:1}):
acknowledges the ownership of the lock by reading \texttt{false} in
Line~\ref{line:wait} (takes $R_I$ units),
performs the work of size $C$ and releases the lock in
Line~\ref{line:small:1} (takes $W$ units);
the parallel section (red, Lines~\ref{line:set}-\ref{line:parallel:2}~and~\ref{line:swap}):
sets \texttt{my\_node.locked} to \texttt{true} (takes $W$), performs the work of size $P$,
performs a non-contended swap (takes $W$) and, possibly,
reads \texttt{true} in Line~\ref{line:wait} (takes $R_I$).
(Here, the swap operation performed after the very first completed critical section is 
counted in the parallel work, as it is executed in the absence of contention.)    
Every other process $i$ operates in the same way: it swaps as early as possible (taking at most $X$),
waits until process $i - 1$ releases the lock, and then performs its critical (blue) and parallel (red) sections.

Depending on the parameters $N$, $C$, $P$, $W$, and $R_I$,
two types of executions are possible.

\begin{figure*}
  \begin{minipage}{0.79\textwidth}
    \begin{subfigure}{\linewidth}
      \resizebox{\textwidth}{!}{
        \begin{tikzpicture}

\node at (-3.8,0.6) {$1$};
\draw (-3.6,0.6) -- (-3.4,0.6);
\draw (-2.4,0.8) -- (-3.4,0.8) -- (-3.4,0.4) -- (-2.4,0.4);
\node at (-2.9,0.6) {$X$};

\draw[fill=blue!10] (-2.4,0.8) -- (-2.4,0.4) -- (0.8,0.4) -- (0.8,0.8) -- cycle;
\draw (-1.6,0.8) -- (-1.6,0.4);
\draw (0.2,0.8) -- (0.2,0.4);
\node at (-2,0.6) {$R_I$};
\node at (-0.7,0.6) {$C$};
\node at (0.5,0.6) {$W$};
\draw [fill=red!10] (12.2,0.4) -- (0.8,0.4) -- (0.8,0.8) -- (12.2,0.8) -- cycle;
\draw (1.4,0.8) -- (1.4,0.4);
\draw (11.6,0.8) -- (11.6,0.4);
\node at (1.1,0.6) {$W$};
\node at (7.5,0.6) {$P$};
\node at (11.9,0.6) {$W$};
\draw [fill=blue!10] (12.2,0.8) -- (12.2,0.4) -- (15.4,0.4) -- (15.4,0.8) -- cycle;
\draw (13,0.8) -- (13,0.4);
\draw (14.8,0.8) -- (14.8,0.4);
\node at (12.6,0.6) {$R_I$};
\node at (13.8,0.6) {$C$};
\node at (15.1,0.6) {$W$};
\draw [fill=red!10] (15.8,0.8) -- (15.4,0.8) -- (15.4,0.4) -- (15.8,0.4);

\node at (-3.8,0) {$2$};
\draw (-3.6,0) -- (-2.4,0);
\draw (-2.4,-0.2) -- (-1.4,-0.2) -- (-1.4,0.2) -- (-2.4,0.2) -- cycle;
\node at (-1.9,0) {$X$};
\draw (-1.4,0) -- (0.8,0);
\draw [fill=blue!10] (0.8,0.2) -- (0.8,-0.2) -- (4,-0.2) -- (4,0.2) -- cycle;
\draw (1.6,0.2) -- (1.6,-0.2);
\draw (3.4,0.2) -- (3.4,-0.2);
\node at (1.2,0) {$R_I$};
\node at (2.5,0) {$C$};
\node at (3.7,0) {$W$};
\draw [fill=red!10] (15.4,0.2) -- (4,0.2) -- (4,-0.2) -- (15.4,-0.2) -- cycle;
\draw (4.6,0.2) -- (4.6,-0.2);
\draw (14.8,0.2) -- (14.8,-0.2);
\node at (4.3,0) {$W$};
\node at (9.7,0) {$P$};
\node at (15.1,0) {$W$};
\draw [fill=blue!10] (15.8,0.2) -- (15.4,0.2) -- (15.4,-0.2) -- (15.8,-0.2);

\node at (-3.8,-0.4) {$\vdots$};

\node at (-3.8,-1.2) {$N$};
\draw (-3.6,-1.2) -- (-0.4,-1.2);
\draw (-0.4,-1) -- (-0.4,-1.4) -- (0.6,-1.4) -- (0.6,-1) -- cycle;
\node at (0.1,-1.2) {$X$};
\draw (0.6,-1.2) -- (7.2,-1.2);
\draw [fill=blue!10] (7.2,-1) -- (7.2,-1.4) -- (10.4,-1.4) -- (10.4,-1) -- cycle;
\draw (8,-1) -- (8,-1.4);
\draw (9.8,-1) -- (9.8,-1.4);
\node at (7.6, -1.2) {$R_I$};
\node at (8.9,-1.2) {$C$};
\node at (10.1,-1.2) {$W$};
\draw [fill=red!10] (15.8,-1) -- (10.4,-1) -- (10.4,-1.4) -- (15.8,-1.4);
\draw (11,-1) -- (11,-1.4);
\node at (10.7,-1.2) {$W$};

\draw [dashed,line width=2] (12.2,1.2) -- (12.2,-1.8);

\end{tikzpicture}
      }
      \captionsetup{width=.8\linewidth}%
      \caption{Case 1: $W + P + W \geq (N - 1) \cdot (R_I + C + W)$.
         Each process enters the critical section without waiting in the queue.}
      \label{fig:execution:1}
    \end{subfigure}
    \begin{subfigure}{\linewidth}
      \resizebox{\textwidth}{!}{
        \begin{tikzpicture}

\node at (-3.8,0.6) {$1$};
\draw (-3.6,0.6) -- (-3.4,0.6);
\draw (-2.4,0.8) -- (-3.4,0.8) -- (-3.4,0.4) -- (-2.4,0.4);
\node at (-2.9,0.6) {$X$};

\draw[fill=blue!10] (-2.4,0.8) -- (-2.4,0.4) -- (0.8,0.4) -- (0.8,0.8) -- cycle;
\draw (-1.6,0.8) -- (-1.6,0.4);
\draw (0.2,0.8) -- (0.2,0.4);
\node at (-2,0.6) {$R_I$};
\node at (-0.7,0.6) {$C$};
\node at (0.5,0.6) {$W$};
\draw [fill=red!10] (8.8,0.4) -- (0.8,0.4) -- (0.8,0.8) -- (8.8,0.8) -- cycle;
\draw (1.4,0.8) -- (1.4,0.4);
\draw (7.4,0.8) -- (7.4,0.4);
\draw (8,0.8) -- (8,0.4);
\node at (1.1,0.6) {$W$};
\node at (4.4,0.6) {$P$};
\node at (7.7,0.6) {$W$};
\node at (8.4,0.6) {$R_I$};
\draw (8.8,0.6) -- (10.4,0.6);
\draw [fill=blue!10] (10.4,0.8) -- (10.4,0.4) -- (13.6,0.4) -- (13.6,0.8) -- cycle;
\draw (11.2,0.8) -- (11.2,0.4);
\draw (13,0.8) -- (13,0.4);
\node at (10.8,0.6) {$R_I$};
\node at (12,0.6) {$C$};
\node at (13.3,0.6) {$W$};
\draw [fill=red!10] (14,0.8) -- (13.6,0.8) -- (13.6,0.4) -- (14,0.4);

\node at (-3.8,0) {$2$};
\draw (-3.6,0) -- (-2.4,0);
\draw (-2.4,-0.2) -- (-1.4,-0.2) -- (-1.4,0.2) -- (-2.4,0.2) -- cycle;
\node at (-1.9,0) {$X$};
\draw (-1.4,0) -- (0.8,0);
\draw [fill=blue!10] (0.8,0.2) -- (0.8,-0.2) -- (4,-0.2) -- (4,0.2) -- cycle;
\draw (1.6,0.2) -- (1.6,-0.2);
\draw (3.4,0.2) -- (3.4,-0.2);
\node at (1.2,0) {$R_I$};
\node at (2.5,0) {$C$};
\node at (3.7,0) {$W$};
\draw [fill=red!10] (12,0.2) -- (4,0.2) -- (4,-0.2) -- (12,-0.2) -- cycle;
\draw (4.6,0.2) -- (4.6,-0.2);
\draw (10.6,0.2) -- (10.6,-0.2);
\draw (11.2,0.2) -- (11.2,-0.2);
\node at (4.3,0) {$W$};
\node at (7.6,0) {$P$};
\node at (10.9,0) {$W$};
\node at (11.6,0) {$R_I$};
\draw (12,0) -- (13.6,0);
\draw [fill=blue!10] (14,0.2) -- (13.6,0.2) -- (13.6,-0.2) -- (14,-0.2);

\node at (-3.8,-0.4) {$\vdots$};

\node at (-3.8,-1.2) {$N$};
\draw (-3.6,-1.2) -- (-0.4,-1.2);
\draw (-0.4,-1) -- (-0.4,-1.4) -- (0.6,-1.4) -- (0.6,-1) -- cycle;
\node at (0.1,-1.2) {$X$};
\draw (0.6,-1.2) -- (7.2,-1.2);
\draw [fill=blue!10] (7.2,-1) -- (7.2,-1.4) -- (10.4,-1.4) -- (10.4,-1) -- cycle;
\draw (8,-1) -- (8,-1.4);
\draw (9.8,-1) -- (9.8,-1.4);
\node at (7.6, -1.2) {$R_I$};
\node at (8.9,-1.2) {$C$};
\node at (10.1,-1.2) {$W$};
\draw [fill=red!10] (14,-1) -- (10.4,-1) -- (10.4,-1.4) -- (14,-1.4);
\draw (11,-1) -- (11,-1.4);
\node at (10.7,-1.2) {$W$};

\draw [dashed,line width=2] (10.4,1.2) -- (10.4,-1.8);

\node at (15.8,0) {};

\end{tikzpicture}
      }
      \captionsetup{width=.8\linewidth}%
      \caption{Case 2: $W + P + W \leq (N - 1) \cdot (R_I + C + W)$.
        Each process waits in the queue before entering the critical section.}
      \label{fig:execution:2}
    \end{subfigure}
    \caption{Examples of executions of the coarse-grained algorithm from
      Figure~\ref{fig:expand:operation}. Blue intervals depict
      critical sections and red intervals depict parallel sections.}
    \label{fig:execution}
  \end{minipage}\hfill
  \begin{minipage}{0.21\textwidth}
    \centering
    \includegraphics[width=\textwidth]{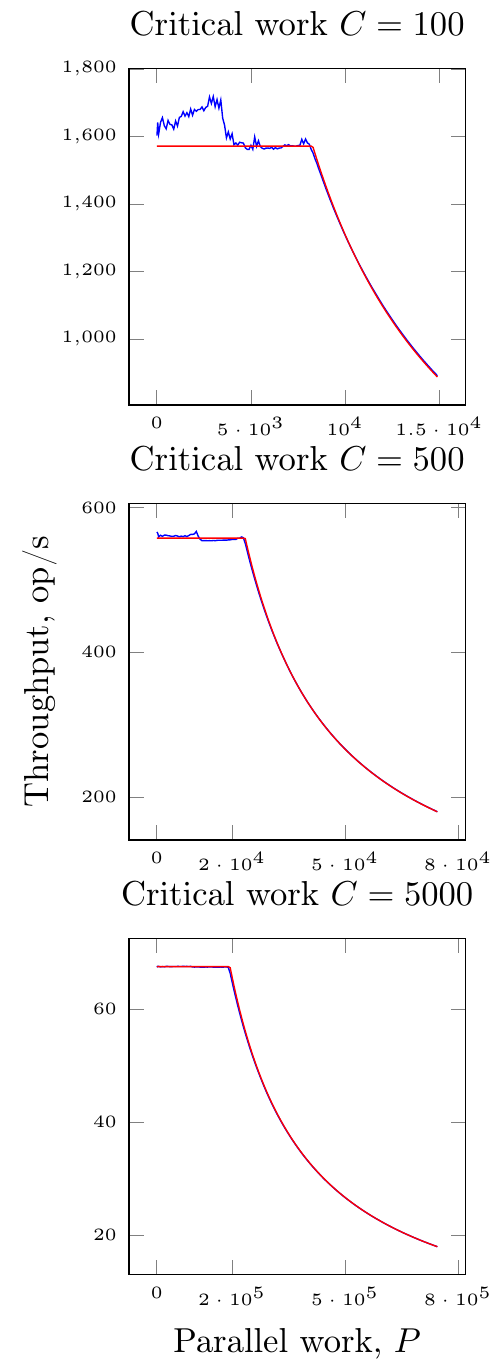}
    \caption{Throughput on $39$ processes for $C \in \{100, 500, 5000\}$}
    \label{fig:plots}
  \end{minipage}
\end{figure*}

In case 1 (Figure~\ref{fig:execution:1}), at the moment when process $1$ finishes its parallel
section, process $N$ already finished its
critical section, i.e.,  $P + 2 \cdot W > (N - 1) \cdot (C + R_I +
W)$.
Therefore, in the steady case, at every moment of time, each process
do not wait and execute either the parallel or critical section,
and the read in Line~\ref{line:wait} cannot return \texttt{true}
because the lock is already released.
Thus, the throughput, measured as the number of 
operations completed in a unit of time, equals to $N \cdot \frac{\alpha}{(P + 2 \cdot W) + (C + R_I + W)}$.

In case 2 (Figure~\ref{fig:execution:2}), before proceeding to the
next operation, process $1$ has to wait until process $N$
completes its critical section from the previous round of operations;
process $2$ waits for process $1$, process $3$ waits for process $2$, etc.  
Thus, there is always some process in the critical section, giving
the throughput of $\frac{\alpha}{C + R_I + W}$.

Therefore, given the number of processes $N$, the sizes $C$ and $P$ of
critical and parallel  sections, the throughput can be calculated as follows:
$$
\begin{cases}
\frac{\alpha}{C + R_I + W} & \text{if } P + 2 \cdot W \leq (N - 1) \cdot (C + R_I + W) \\
\frac{\alpha \cdot N}{(P + 2 \cdot W) + (C + R_I + W))} & \text{otherwise}
\end{cases}
$$

\setlength{\belowcaptionskip}{-5pt}
\setlength{\abovecaptionskip}{10pt}

\section{Experiments}
For our measurements, we used a server with four 10-core Intel Xeon E7-4870
chips of 2.4 GHz (yielding 40 hardware processes in total),
running Ubuntu Linux kernel v3.13.0-66-generic.
We compiled the code with MinGW GCC 5.2.0 (with -O0 flag to avoid
compiler optimizations, such as function inlining,
that can screw up our benchmarking environment).
The code is available at \\ \url{https://github.com/Aksenov239/complexity-lock-with-libslock}.

We considered the following experimental settings: 
the number of processes $N \in \{5, 10, 20, 30, 39\}$;
the size of the critical section $C \in \{100, 500, 1000, 5000, 10000\}$; 
and the multiplier $x \in [1, 150]$ (we choose all integer values)
that determined the size of the parallel section $P = x \cdot C$.
For each setting, we measured the throughput for 10 seconds.
Our experimental evaluation gives $\alpha \approx 3.5 \cdot 10^5$,
$W \approx 40$, and $R_I \approx 80$.
The ratio between $W$ and $R_I$ correlates with the experimental results provided
by David et al.~\cite{david2013everything}.

In Figure~\ref{fig:plots} we show our experimental results for three settings with
$N = 39$ and $C \in \{100, 500, 5000\}$ (blue curves) compared with our
theoretical prediction (red curves).
%
%
The two curves match very closely, except for the case of small $C$ and $P$
where our predicted throughput underestimates the real one.
%
%
We relate this to the fact that we oversimplified
the abstract machine: any write induces
the work of constant size $W$, regardless of
the relative location of the cache line
with respect to the process.
For small $C$ and $P$
two processes from the same socket are more
likely to take the lock one after the other and,
thus, on average, a write might induce less work than $W$,
and, consequently, the throughput can be higher than
predicted.

\section{Comparison with Prior Work}
In this work, we proposed a very simple, not to say simplistic, analytical framework intended
to predict the performance of a class of \emph{lock-based} algorithms.
A more involved analysis has been earlier proposed by Atalar et al.~\cite{tsigas15}
for a similar class of \emph{lock-free} concurrent data structures (Figure~\ref{fig:lock-free}).
There the concurrent processes alternate the constant size parallel work with constant-size
critical work and synchronize critical operations on the shared data using
read and \texttt{compare\&set} operations on a decdicated \emph{access point}.
\begin{figure}
\begin{lstlisting}
operation():
  parallel_work()
  while !success do                             
    current $\leftarrow$ read(AP)                 |\label{line:retry:start}|
    new $\leftarrow$ critical_work(current)
    success $\leftarrow$ compare&set(AP, current, new)    |\label{line:retry:end}|
\end{lstlisting}
\caption{The lock-free operation}
\label{fig:lock-free}
\end{figure}

By adapting the code to our notations, we get:
\begin{lstlisting}
operation():
  for i in 1..P:
    nop
  while !success:
    current = read(AP)
    for i in 1..C-1:
      nop
    it++
    success = compare&set (AP, current, (thread_id, it))
\end{lstlisting}

In order to have the critical work of size $C$ we had to have the critical loop of size $C - 1$,
because in each iteration of the loop we increment the variable \texttt{i} and after the loop
we increment the thread local variable \texttt{it}.

We argue that the two approaches, ours and by Atalar et al.~\cite{tsigas15}
though seemingly quite similar, bear some important differences.
In particular, these differences, do not allow us to treat our analytical framework as
a special case of that in~\cite{tsigas15}.

Note that we do not consider here the more general analysis in a later paper by Atalar~\cite{tsigas16}
in which the amounts of parallel and critical work are treated as random variables obeying specific distributions.
The analysis in~\cite{tsigas16} is a probabilistic generalization of that in~\cite{tsigas15}.
Therefore, it appears that, for the sake of comparison, we can focus on the deterministic framework of~\cite{tsigas15}.

Two types of conflicts happen in the described lock-free algorithms:
\begin{itemize}
\item \emph{logical conflicts}~--- the unsuccessful retry
  Lines~\ref{line:retry:start}-\ref{line:retry:end}, i.e.,
  a ``fast'' process succeeds in updating the \emph{access point} variable \texttt{AP}, causing ``slower'' process
  to fail in their \texttt{compare\&set} operations;
\item \emph{hardware conflicts} relates here to the serialization of concurrent reads at Line~\ref{line:retry:start}
  and \texttt{compare\&set} at Line~\ref{line:retry:end} on \texttt{AP}.
\end{itemize}

At the same time, our lock-based algorithms are subject only to hardware conflicts on \texttt{head} variable
(Figure~\ref{fig:expand:operation} Line~\ref{line:swap}).

At first, let us look on the two types of algorithms from the high-level point of view.
Lock-based algorithms are \emph{conservative} in the sense that the critical section is performed
only when the lock is taken and the actions of the critical section always ``take place''.
In particular, this kind of algorithms is only subject to hardware conflicts.

In contrast, lock-free algorithms are \emph{speculative}: a critical section can be performed several times
before it succeeds and only the actions of the successful instance are effective.
In analysing these algorithms, we should account for both logical and hardware conflicts.

Under high contention, speculative data structures peform worse than conservative ones due to the orverwhelming
number of retries critical sections.
Intuitively, this suggests that we should use different analyses to reason about the throughput
of these two classes of algorithms.

In what follows, we suppose that $P$ and $C$ exceed the cost of the \texttt{swap} operation.
Such condition greatly simplifies the analysis for lock-based data structures since we do not have to deal with
hardware conflicts.

Under high contention, i.e., when $P$ is comparatively small, $P << (n
- 1) \cdot C$,
we use the special properties of the lock-based algorithms: with
these parameters there is always some process
in the critical section and, consequently, this allows us to easily
evaluate the resulting throughput.

In contrast, the analysis of the performance of lock-free algorithms
under high contention in~\cite{tsigas15} is considerably more
involved, due to the intrinsic interleaving of hardware and logical conflicts.

When contention is small and conflicts are unlikely,
the two analyses for fixed $P$ and $C$ should coincide. 
Both of them provide us with the throughput approximately equal to $\frac{\alpha N}{P + C}$ where $\alpha$ is some constant
and $N$ is the number of threads.

To summarize, in the case of CLH Lock the analysis for lock-based algorithms
coincide with the analysis for lock-free programs for the settings with small contention, while
in other settings our analysis is much simpler due to the special properties of the lock-based algorithms.

Furthermore, we consider the MESI cache-coherence protocol~\cite{david2013everything}.
Our analysis is further simplified by assuming that
writes take the same time no matter in which state a cache line is:
there are evidences that this is indeed the case for our machine.
However, the situation might get more complicated for other machines
in which, e.g., the write complexity depends on the cache state, which
might result in a more complicated analysis.

If the CLH lock in conservative programs is replaced with a lock of
another type, e.g., test\&test\&set,
ticket, spin lock, MCS, etc., the analysis becomes somewhat more complicated 
but it still shares the part when $P \geq (n - 1) \cdot C$.

For example, suppose that we replace CLH lock with spin lock:
\begin{lstlisting}
operation():
  while !success:                
    success = compare&set(locked, 0, 1)  
  for i in 1..C:                 
    nop                          
  locked = 0
  for i in 1..P:
    nop
\end{lstlisting}

Here we have hardware conflicts not only on compare\&set, but also on the write \texttt{locked = 0}.
These conflicts are not considered by the analysis of lock-free
programs and, thus, there should be a different analysis for
the coarse-grained programs with spin lock.

To summarize, the two analyses, though designed using similar
arguments, are distinct.

\section{Conclusion} 
In this short note, we showed that a simple theoretical analysis
may quite accurately predict the throughput of data
structures implemented using coarse-grained synchronization.
For the moment, our analysis is restricted to
algorithms using CLH-based locking in systems
obeying the uniform scheduler.
In upcoming work, we intend extend the analysis
to more realistic algorithm designs, lock implementations and architectures.

\section{Acknowledgements}
This research is partially supported by
European Research Council (ERC-2012-StG-308246) and the Franco-German DFG-ANR Project DISCMAT (14-CE35-0010-02).

\bibliography{references}

\end{document}